\documentclass[aps,prl,twocolumn,10pt,showpacs,superscriptaddress,amsmath,amssymb,nobibnotes]{revtex4-1}

\usepackage{graphicx}
\usepackage{amsfonts}    
\usepackage{graphicx}   
\usepackage{verbatim}   
\usepackage{color}      
\usepackage{subfigure}  
\usepackage{hyperref}   
\usepackage{natbib}
\usepackage{booktabs}
\usepackage{threeparttable}
\usepackage{dcolumn}
\usepackage{multirow}
\usepackage{epsfig}

\begin{document}
\title{Direct Measurement of the Two-dimensional Spatial Quantum Wavefunction via Strong Measurements}
\author{Chen-Rui Zhang}\email{This authors contribute equally to this work}
\author{Meng-Jun Hu}\email{This authors contribute equally to this work}
\author{Zhi-Bo Hou}
\author{Jun-Feng Tang}
\author{Jie Zhu}
\author{Guo-Yong Xiang}\email{gyxiang@ustc.edu.cn}
\author{Chuan-Feng Li}
\author{Guang-Can Guo}
\author{Yong-Sheng Zhang}\email{yshzhang@ustc.edu.cn}
\affiliation{Laboratory of Quantum Information, University of Science and Technology of China, Hefei 230026, P. R. China}
\affiliation{CAS Center for Excellence in Quantum Information and Quantum Physics, University of Science and Technology of China, Hefei 230026, P. R. China}

\date{\today}

\begin{abstract}
Wavefunction is the foundation of quantum theory, which is assumed to give a complete description of a quantum system. For a long time, wavefunction is introduced as an abstract element of the theory and there lacks effective ways to measure it directly.
The situation, however, is somewhat changed when Lundeen {\it et al}. reported the direct measurement of the quantum wavefunction via weak measurements, which gives the wavefunction a clearly operational definition [Nature {\bf 474}, 188 (2011)]. The weak measurement method requires sequential measurements of conjugate observables position and momentum with the position measurement is weak enough. 
Surprisingly, the recent research by Vallone and Dequal shows that performing sequential strong measurements realizes the same target, in which case no approximation has to be made compared to the case of weak measurements[Phys. Rev. Lett. {\bf 116}, 040502 (2016)]. Here we experimentally report the direct measurement of the two-dimensional transverse wavefunction of photons via strong measurements for the first time, which implies that an accurate and clear operational definition can be given to wavefunction. We have measured the Gaussian and Laguerre-Gaussian of $l=1$ spatial wavefunctions of photons with R-square are $0.97$ and $0.93$ respectively.
As a potentially important application, we show that the direct measurement of two-dimensional wavefunction provides an alternative way to realize digital holography of three-dimensional objects. The results presented here will not only deepen our understanding of abstract wavefunction but also have significant applications in quantum information processing and quantum imaging.

\end{abstract}

\maketitle
As the core of quantum theory, the quantum system is described by the quantum state $|\psi\rangle$ or wavefunction $\psi$ \cite{w1,w2,w3}. According to Born rule, the modulus square of wavefunction predict the probabilities of all kinds of measurement outcomes \cite{Born}.  Since the beginning, the mystery wavefunction confuses generations of physicists, most of them eventually give away and take the pragmatic attitude following the philosophy of Copenhagen `Shut up and calculate!' \cite{Mermin}. The question about the reality of wavefunction, however, never fades away \cite{EPR, Bohm, Dewitt, wave}. Debates on whether or not wavefunction represents subjective knowledge or information about some aspect of reality \cite{sub1, sub2, sub3} or corresponds directly to objective reality \cite{ob1, ob2, ob3} continue with foreseeable future.

From the point of view of practical applications, on the other hand, an operational definition of wavefunction, i.e., how reconstruct the wavefunction of the quantum system is required. Unfortunately, the unknown state of a single quantum is impossible to be determined due to the facts that measurement inevitably disturb the system and no-cloning theorem \cite{zurek}.
For an ensemble of quantum states undergo the same preparation process with individual described by the same wavefunction $\psi$, the wavefunction $\psi$ can be reconstructed by quantum state tomography (QST) \cite{qst1, qst2, qst3}, which requires measuring a large number of complete sets of noncommuting observables of system. Construction of $\psi$ by QST needs complicated computations with the increasing dimensions of measured system. Indeed the transverse wavefunction of single photons has never been reported via QST method.

Significant progress was made by Lundeen {\it et al}. who reported the realization of direct measurement of transverse wavefunction of photons \cite{Lundeen} based on the weak measurement \cite{weak1, weak2, weak3}. This method, which called direct weak tomography (DWT) \cite{S}, provides a clearly operational definition of wavefunction by measuring it directly and avoids the complicated measurements and calculations in the case of QST. The "direct measurement" here in DWT refers to that the real and imaginary part of the wavefunction being probed at a spatial point is proportional to the real and imaginary part of weak value at that point, therefore it can be directly obtained based on the measurement probabilities. The DWT method of measuring wavefunction rests on the sequential measurement of conjugate observables of system, i.e., position and momentum with the first measurement of position $\pi_{x}\equiv|x\rangle\langle x|$ is weak enough followed by a strong measurement of momentum \cite{S}. Suppose that the initial state of particles being measured is $|\psi\rangle$, the weak value of $\pi_{x}$ given that momentum $P=p$ is
\begin{equation}
\langle \pi_{x}\rangle_{w}=\dfrac{\langle p|x\rangle\langle x|\psi\rangle}{\langle p|\psi\rangle}=\dfrac{e^{ipx/\hbar}}{\phi(p)}\psi(x).
\end{equation}
Choosing $p=0$, the wavefunction of particles is thus determined by
\begin{equation}
\psi(x)=\phi(0)\langle\pi_{x}\rangle_{w}
\end{equation}
with $\phi(0)$ is a constant. The real and imaginary part of weak value $\langle\pi_{x}\rangle_{w}$ can be obtained by performing proper  complementary observables measurement on the pointer \cite{Josza}. Scanning step by step along $x$ direction, the wavefunction distribution $\psi(x)$ can be achieved.

\begin{figure}[tbp]
\centering
\includegraphics[scale=0.35]{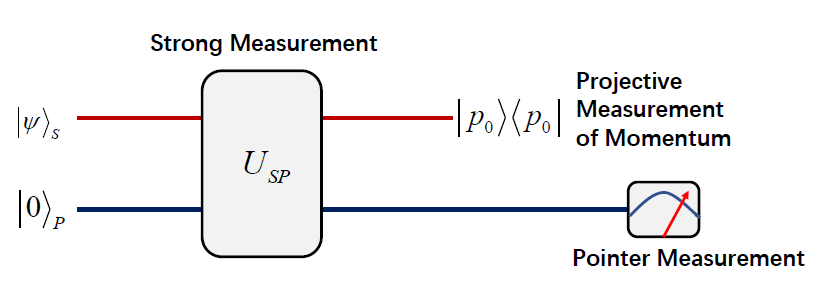}
\caption{{\bf Scheme of direct strong tomography(DST).} In the case of two-dimensional transverse wavefunction measurement, $|\psi\rangle_{S}=\sum^{d}_{x,y=1}\psi(x,y)|x,y\rangle$, $|0\rangle_{P}=(|H\rangle+|V\rangle)/\sqrt{2}$, $\hat{U}_{SP}=e^{-i\theta|x,y\rangle\langle x,y|\otimes\hat{\sigma}_{y}}$ with $\theta=\pi/2$ represent strong position measurement, $|p_{0}\rangle=\sum_{x,y}|x,y\rangle/\sqrt{d}$ is the zero transverse momentum state.}
\label{0}
\end{figure}

The DWT method has been widely applied in the measurement of photonic state \cite{app1, app2, app3, app4} since the report of wavefunction measurement. For a period of time, people believed that weak measurement in DWT is the key to realize direct tomography of quantum state. Surprisingly, the recent work of Vallone and Dequal shows that it is not the case and strong measurement gives the better direct measurement of wavefunction \cite{S}. The strong measurement method, which called direct strong tomography (DST), has the same framework with that in DWT but with the measurement of position be strong, i.e., the interaction between system and pointer is strong. The interaction strength of position measurement in DST actually can be arbitrary and thus DWT is only the weak limit case of DST. The DST method has been applied in discrete system and shown better accuracy and precision in state reconstruction than DWT \cite{matter,Density matrix}.

In this work, we experimentally report the direct measurement of the two-dimensional spatial wavefunction of photons based on the DST method. For photons travelling along $z$ direction, we directly measure its transverse wavefunction $\psi(x,y)$ in the $x-y$ plane. In particular, wavefunction of photons carrying orbital angular momentum is measured. As a potentially important application, we discuss the digital holography of three-dimensional objects based on the direct measurement of two-dimensional spatial wavefunction of light.


\begin{figure}[htbp]
\centering
\includegraphics[scale=0.5]{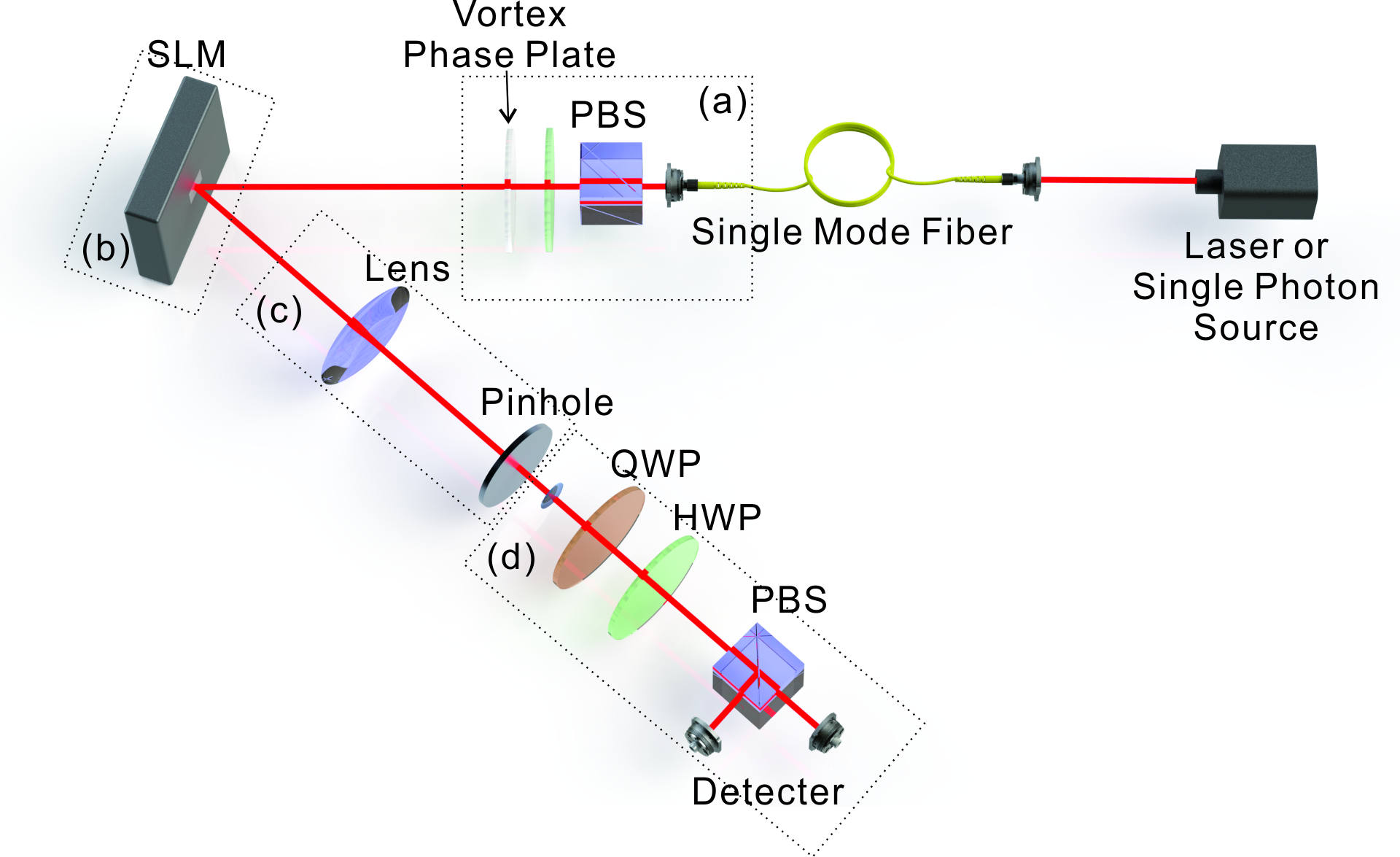}
\caption{{\bf Experiment setup for measurement of transverse wavefunction via direct strong tomography.} The setup consists of four modules: (a) preparation of initial state of photons as $\int\psi(x,y)|x,y\rangle\otimes|0\rangle_{P}dxdy$ with $\psi(x,y)$ be Gaussian or Laguerre-Gaussian of $l=1$ and $|0\rangle_{P}=(|H\rangle+|V\rangle)/\sqrt{2}$; (b) strong measurement of $|x,y\rangle\langle x,y|$ is performed via a phase-only spatial light modulator (Hamamatsu X10468) with $10\times 10$ pixel block for each scanning step and each pixel is about $12.5\mathrm{\mu m}\times 12.5\mathrm{\mu m}$; (c) post-selection of system into zero transverse momentum via a Fourier transform lens (Thorlabs AC508-400-B-ML) with focal length $f=40\mathrm{cm}$ and a pinhole with diameter of $50\mathrm{\mu m}$; (d) pointer measurement with a polarization analyser to directly extract real and imaginary part of wavefunction. PBS: Polarizing beam splitter; SLM: Spatial light modulator; HWP: Half wave plate; QWP: Quarter wave plate. }
\label{1}
\end{figure}

The scheme of DST is shown in Fig. \ref{0}. Consider particles travelling along $z$ direction with initial state of system be $|\psi\rangle_{S}$, in the position representation $|\psi\rangle_{S}=\int\psi(x,y)|x,y\rangle dxdy$ with $\psi(x,y)$ is the two-dimensional transverse wavefunction to be measured. It is more appropriate to recast in the discrete form as $|\psi\rangle_{S}=\sum^{d}_{x,y=1}\psi(x,y)|x,y\rangle$ in practice due to the limited precision of measurement apparatus. The pointer is chosen as polarization qubit with initial state $|0\rangle_{P}=(|H\rangle+|V\rangle)/\sqrt{2}$ in our case. Here $|H\rangle, |V\rangle$ represent horizontal and vertical polarization of photons respectively. The composite state of the system and the pointer, after interaction, evolves into
\begin{equation}
|\Psi\rangle_{SP}=\hat{U}_{SP}|\psi\rangle_{S}\otimes|0\rangle_{P}=e^{-i\theta|x,y\rangle\langle x,y|\otimes\hat{\sigma}_{y}}|\psi\rangle_{S}\otimes|0\rangle_{P},
\end{equation}
where $\theta$ represents the strength of coupling and $\hat{\sigma}_{y}$ is a Pauli operator defined in the basis $\lbrace|0\rangle, |1\rangle\rbrace$.
In the case of strong position measurement $\theta=\pi/2$, $\hat{U}_{SP}=\hat{I}-|x,y\rangle\langle x,y|\otimes(\hat{I}+i\hat{\sigma}_{y})$, we thus have
\begin{equation}
|\Psi\rangle_{SP}=|\psi\rangle_{S}\otimes|0\rangle_{P}-\sqrt{2}\psi(x,y)|x,y\rangle\otimes|-\rangle_{P}
\end{equation}
with $|-\rangle_{P}=(|0\rangle_{P}-|1\rangle_{P})/\sqrt{2}$ and $|1\rangle_{P}=(|H\rangle-|V\rangle)/\sqrt{2}$. After strong measurement of position, we perform projective measurement of momentum with resulting state $|p_{0}\rangle=\sum_{x,y}|x,y\rangle/\sqrt{d}$ stands for zero transverse momentum.  The state of pointer, given that post-selected momentum state is $|p_{0}\rangle$, becomes (unnormalized)
\begin{equation}
|\varphi\rangle_{P}=\langle p_{0}|\Psi\rangle_{SP}=\dfrac{1}{\sqrt{d}}[\tilde{\psi}|0\rangle_{p}-\sqrt{2}\psi(x,y)|-\rangle_{P}],
\end{equation}
where $\tilde{\psi}=\sum_{x,y}\psi(x,y)$. The phase of $\tilde{\psi}$ is a global phase for every $\psi(x,y)$, which can be chosen such that $\tilde{\psi}$ is real valued and positive.  The real and imaginary part of wavefunction $\psi(x,y)$ can be obtained by performing proper measurements on the pointer, specifically we have
\begin{equation}
\begin{split}
\mathrm{Re}[\psi(x,y)]&=\dfrac{d}{2\tilde{\psi}}(P_{|+\rangle}+2P_{|1\rangle}-P_{|-\rangle}),  \\
\mathrm{Im}[\psi(x,y)]&=\dfrac{d}{2\tilde{\psi}}(P_{|L\rangle}-P_{|R\rangle}),
\end{split}
\end{equation}
where $P_{|\pm\rangle}=|\langle\pm|\varphi\rangle|^{2}, P_{|L\rangle}=|\langle L|\varphi\rangle|^{2}, P_{|R\rangle}=|\langle R|\varphi\rangle|^{2}$ and $P_{|1\rangle}=|\langle 1|\varphi\rangle|^{2}$ with $|\pm\rangle=(|0\rangle\pm |1\rangle)/\sqrt{2}, |L\rangle=(|0\rangle+i|1\rangle)/\sqrt{2}$ and $|R\rangle=(|0\rangle-i|1\rangle)/\sqrt{2}$. It should be noted that the above equations are accurate while in the DWT method the first order approximation is taken.

Our experiment setup for realizing direct wavefunction measurement of photons via DST is shown in Fig. \ref{1}. The setup consists of four stages: preparation of the initial states of photons, strong measurement of transverse position of photons, post-selection of photons with zero transverse momentum, and readout of measurement.

\begin{figure}[tbp]
\centering
\includegraphics[scale=0.08]{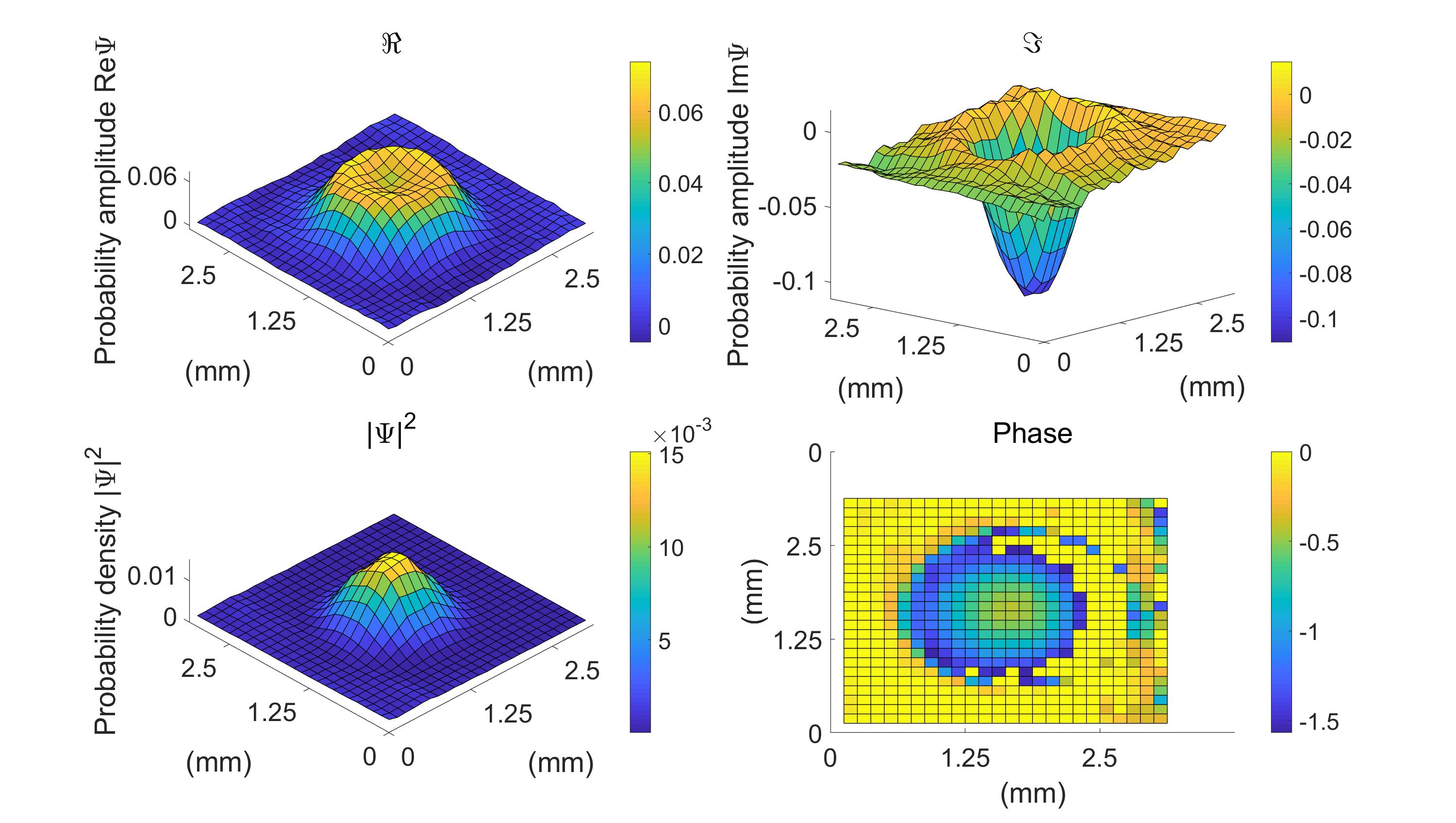}
\caption{{\bf Gaussian spatial wavefunction}. The real and imaginary part of wavefunction are shown in the upper row while the corresponding probability density and phase distribution are shown in  the lower row. The R-square between the measured data of probability density and the ideal Gaussian distribution is $0.97$.   }
\label{2}
\end{figure}

The stream of photons with centre wavelength of $808\mathrm{nm}$ are produced in one of two ways, either by a laser beam or by generating heralded single photons through spontaneous parametric down-conversion (SPDC) via pumping a type-II Beta Barium Borate (BBO) crystal.
Photons are then coupled into a single mode fiber (SMF) and collimated after emitting from the SMF, with the transverse mode is approximately Gaussian. The initial state of pointer i.e., the polarization of photons is prepared via a polarizing beam splitter (PBS) and a half wave plate (HWP) rotated by $\pi/8$ radian. Photons that carry orbital angular momentum $l=1$ are produced by adding a vortex phase plate after the HWP in our experiment. Photons with higher angular momentum state can be realized via spatial light modulator (SLM).
This completes the preparation of the initial states of photons as $\int\psi(x,y)|x,y\rangle\otimes|0\rangle_{P}dxdy$ with $|0\rangle_{P}=(|H\rangle+|V\rangle)/\sqrt{2}$ and the two-dimensional transverse wavefunction $\psi(x,y)$ be Gaussian or Laguerre-Gaussian of $l=1$.

We implement strong measurement of system by using a phase-only SLM, which rotates the polarization of photons from state $|0\rangle_{P}$ into $|1\rangle_{P}=(|H\rangle-|V\rangle)/\sqrt{2}$ at the block area with $10\times 10$ pixels and each pixel is about $12.5\mathrm{\mu m}\times 12.5\mathrm{\mu m}$. This realizes von Neumann-type interaction Hamiltonian $\hat{H}=\theta|x,y\rangle\langle x,y|\otimes\hat{\sigma}_{y}$ with $\theta=\pi/2$ and correspondingly unitary transformation $\hat{U}=e^{-i\hat{H}}$.
The post-selection of system into zero transverse momentum is then realized via a Fourier transform lens with focal length $f=40\mathrm{cm}$ and a pinhole with diameter of $50\mathrm{\mu m}$.
The post-selected photons are subsequently sent into a polarization analyser, which consists of a HWP, a quarter wave plate (QWP), a PBS and two detectors, for final readout. If we regard the polarization as a spin-1/2 system, then the real part and imaginary part of wavefunction $\psi(x,y)$ are proportional to expectation values of Pauli observables according to Eq. (6) as
\begin{equation}
\begin{split}
&\mathrm{Re}[\psi(x,y)]\propto \langle\varphi|(\hat{\sigma}_{x}+2|1\rangle\langle 1|)|\varphi\rangle  \\
&\mathrm{Im}[\psi(x,y)]\propto \langle\varphi|\hat{\sigma}_{y}|\varphi\rangle,
\end{split}
\end{equation}
where $|\varphi\rangle$ represent the polarization state of the post-selected photons, and $\hat{\sigma}_{x}\equiv P_{|+\rangle}-P_{|-\rangle}$ and $\hat{\sigma}_{y}\equiv P_{|L\rangle}-P_{|R\rangle}$ define the Pauli $x$ and $y$ matrices respectively.

\begin{figure}[tbp]
\centering
\includegraphics[scale=0.08]{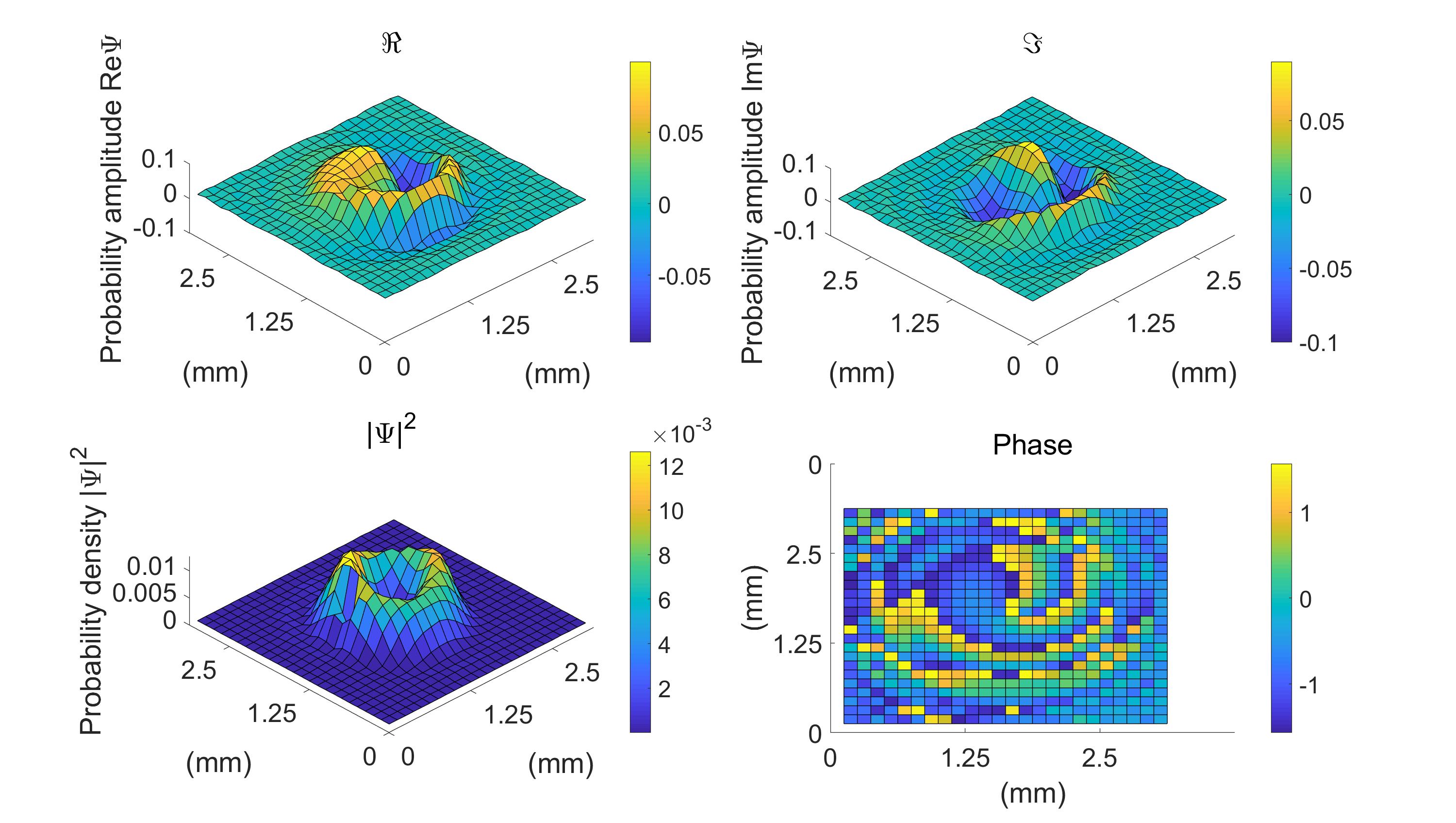}
\caption{{\bf Laguerre-Gaussian spatial wavefunction of $l=1$.} The real and imaginary part of wavefunction are shown in the upper row while the corresponding probability density and phase distribution are shown in the lower row. The R-square between the measured data of probability density and the ideal Laguerre-Gaussian distribution of $l=1$ is $0.93$.}
\label{3}
\end{figure}

In order to obtain full distribution of wavefunction $\psi(x,y)$, we perform scanning on the $x-y$ plane of the SLM step by step with elementary area of $10\times 10$ pixels. The measurement of real and imaginary part of the wavefunction directly gives the probability distribution $|\psi|^{2}=\mathrm{Re}^{2}\psi+\mathrm{Im}^{2}\psi$ and phase distribution $\vartheta(x,y)=\mathrm{arctan}(\mathrm{Re}\psi/\mathrm{Im}\psi)$.
By not inserting or inserting a vortex phase plate, we have measured the Gaussian spatial wavefunction of photons and wavefunction of photons carrying orbital angular momentum of $l=1$ as shown in Fig. \ref{2} and Fig. \ref{3} respectively. The figures are obtained via fitting of measured data and the quality of fit measured by R-square between the measured data of probability density and the ideal distribution are approximately $0.97$ and $0.93$ for Gaussian and Laguerre-Gaussian of $l=1$ case respectively.

\begin{figure}[tbp]
\centering
\includegraphics[scale=0.26]{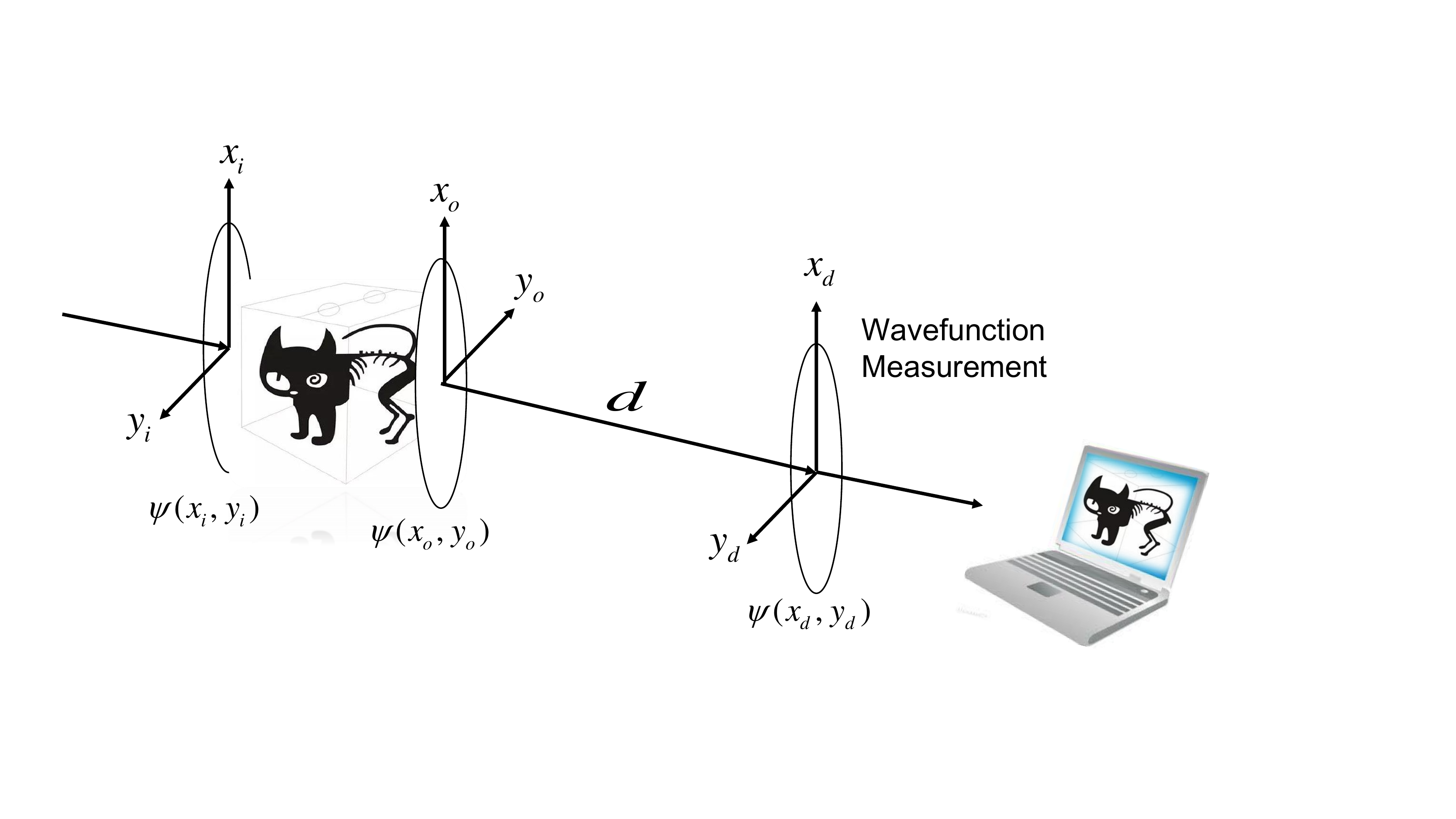}
\caption{{\bf Schematic of quantum digital holography.} The wavefunction $\psi(x_{o},y_{o})$ in the object plane $(x_{o},y_{o})$ can be inferred based on the measured wavefunction $\psi(x_{d},y_{d})$ in the imagine plane $(x_{d},y_{d})$ via Fast Fourier transformation algorithm according to Eq. (10). The object is reconstructed by comparing the $\psi(x_{o},y_{o})$ with the input known wavefunction $\psi(x_{i},y_{i})$ via computer program.}
\label{4}
\end{figure}

As a potentially important application, the direct measurement of two-dimensional spatial wavefunction of photons provides an  alternative way to realize the digital holography of three-dimensional objects. In conventional digital holography, the object light has to be interfere with reference light \cite{holo1,holo2}. In order to obtain the complex distribution of the object light field, a phase-shifted interferometry is used, i.e., the phase of object light field is inferred by modulating the phase of reference light. In contrast, the direct wavefunction measurement method provides a direct way to obtain the complex distribution of object light field without the help of reference light.

Suppose that photons with known wavefunction $\psi(x_{i},y_{i})$ be incident to an object, after passing through object the reflected or transmitted wavefunction of photons becomes $\psi(x_{o},y_{o})$ as shown in Fig. \ref{4}. The object can be reconstructed by comparing the measured $\psi(x_{o},y_{o})$ with known $\psi(x_{i},y_{i})$ via computer program. In practice what we measure is the wavefucntion in the imagine plane $(x_{d},y_{d})$ and it is determined by object wavefunction $\psi(x_{o},y_{o})$ via
\begin{equation}
 \psi(x_{d},y_{d})=\int K(x_{d},y_{d};x_{o},y_{o})\psi(x_{o},y_{o})dx_{o}dy_{o}
 \end{equation}
where $K(x_{d},y_{d};x_{o},y_{o})$ is the Feynman propagator
\begin{equation}
K(x_{d},y_{d};x_{o},y_{o})=\dfrac{1}{i\lambda}\dfrac{e^{ik\sqrt{(x_{d}-x_{o})^{2}+(y_{d}-y_{o})^{2}+d^{2}}}}{\sqrt{(x_{d}-x_{o})^{2}+(y_{d}-y_{o})^{2}+d^{2}}}
\end{equation}
with $\lambda, k$ are the wavelength and the number of wavevector of photons respectively and $d$ is the distance from the object plane to imagine plane.

When $d\gg 1$, the paraxial condition is satisfied, the object's wavefunction is given by the inverse transformation of Eq. (7)
\begin{equation}
 \psi(x_{o},y_{o})=\int L(x_{o},y_{o};x_{d},y_{d})\psi(x_{d},y_{d})dx_{d}dy_{d}
 \end{equation}
with
\begin{equation}
L(x_{o},y_{o};x_{d},y_{d})=\dfrac{e^{-ikd}}{-i\lambda d}\mathrm{exp}\lbrace\dfrac{-ik[(x_{d}-x_{o})^{2}+(y_{d}-y_{o})^{2}]}{2d}\rbrace.
\end{equation}
With the help of Fast Fourier transformation algorithm, we can obtain the object wavefunction $\psi(x_{o},y_{o})$ from the measured wavefunction $\psi(x_{d},y_{d})$ in the imagine plane and then reconstruct object by comparing with the input known wavefunction $\psi(x_{o}, y_{o})$ via computer program.

In conclusion, we have experimentally demonstrated the measurement of two-dimensional spatial wavefunction of photons via the method of direct strong tomography, which provides an accurately and clearly operational definition of wavefunction. The realization of two-dimensional spatial wavefunction offers a quantum way to realizing digital holography of three-dimensional objects. When combined with compressive algorithm \cite{compressive} and scan-free approach \cite{app4}, the direct wavefunction measurement provides the possible real-time imagining of objects. The results not only deepen our understanding of abstract wavefunction but also may have significant applications in quantum information processing and quantum digital holography.

This work was supported by the National Natural Science Foundation of China (Grant Nos. 11574291, 11774334, 61327901, 11774335, 61275122, 11674306 and 61590932), the Strategic Priority Research Program (B) of the Chinese Academy of Sciences (No. XDB01030200), the National Key Research and Development Program of China (No. 2016YFA0301300, No. 2016YFA0301700 and No. 2017YFA0304100), Key Research Program of Frontier Science, CAS (No. QYZDY-SSW-SLH003) and Anhui Initiative in Quantum Information Technologies.


\begin{thebibliography}{99}
\bibitem{w1} C. Cohen-Tannoudji, B. Diu, and F. Laloe, Quantum Mechanics Vol. 1, p19 (Wiley-Interscience, 2006).

\bibitem{w2} L. D. Landau and E. M. Lifshitz, Course of Theoretical Physics Vol. 3, Quantum Mechanics:Non-Relativistic Theory 3rd edn, p6 (Pergamon, 1989).

\bibitem{w3} P. A. M. Dirac, The Principles of Quantum Mechanics 4rd edn (Oxford University, 1958).


\bibitem{Born} M. Born, Quantenmechanik der Sto$\ss$vorg\"{a}nge. Zeit. Phys. {\bf 38}, 803-827 (1926).

\bibitem{Mermin} N. D. Mermin, Could Feynman Have Said This? Physics Today {\bf 57}, 10 (2004).

\bibitem{EPR} A. Einstein, B. Podolsky, and N. Rosen, Can Quantum-Mechanical Description of Physical Reality Be Considered Complete? Phys. Rev. {\bf 47}, 777 (1935).

\bibitem{Bohm} D. Bohm, A Suggested Interpretation of the Quantum Theory in Terms of ``Hidden" Variables. I. Phys. Rev. {\bf 85}, 166 (1952).

\bibitem{Dewitt} B. S. DeWitt, Quantum mechanics and reality. Physics Today {\bf 23}, 30 (1970).

\bibitem{wave} Y. Aharonov, J. Anandan, and L. Vaidman, Meaning of the wave function. Phys. Rev. A {\bf 47}, 4616 (1993).

\bibitem{sub1} L. E. Ballentine, The statistical interpretation of quantum mechanics. Rev. Mod. Phys. {\bf 42}, 358-381 (1970).

\bibitem{sub2} C. M. Caves, C. A. Fuchs, and R. Schack, Quantum probabilities as bayesian probabilities. Phys. Rev. A {\bf 65}, 022305 (2002).

\bibitem{sub3} R. W. Spekkens, Evidence for the epistemic view of quantum states: A toy theory. Phys. Rev. A {\bf 75}, 032110 (2007).

\bibitem{ob1} M. F. Pusey, J. Barrett, and T. Rudolph, On the reality of quantum state. Nature Phys. {\bf 8}, 476-479 (2012).

\bibitem{ob2} M. Ringbauer, B. Duffus, C. Branciard, E. G. Cavalcanti, A. G. White, and A. Fedrizzi, Measurements on the reality of the wavefunction. Nature Phys. {\bf 11}, 249-254 (2015).

\bibitem{ob3} R. Colbeck and R. Renner, Is a System's Wave Function in One-to-One Correspondence with Its Elements of Reality? Phys. Rev. Lett. {\bf 108}, 150402 (2012).

\bibitem{zurek} W. K. Wootters and W. H. Zurek, A single quantum cannot be cloned. Nature {\bf 299}, 802-803 (1982).

\bibitem{qst1} D. T. Smithey, M. Beck, M. G. Raymer, and A. Faridani, Measurement of the Wigner distribution and the density matrix of a light mode using optical homodyne tomography: Application to squeezed states and the vacuum. Phys. Rev. Lett. {\bf 70}, 1244 (1993).

\bibitem{qst2} G. Breitenbach, S. Schiller, and J. Mlynek, Measurement of the quantum states of squeezed light. Nature {\bf 387}, 471–475 (1997).

\bibitem{qst3} D. F. V. James, P. G. Kwiat, W. J. Munro, and A. G. White, Measurements of qubits. Phys. Rev. A {\bf 64}, 052312 (2001).

\bibitem{Lundeen} J. S. Lundeen, B. Sutherland, A. Patel, C. Stewart, and C. Bamber, Direct measurement of the quantum wavefunction. Nature {\bf 474}, 188-191 (2011).

\bibitem{weak1} Y. Aharonov, D. Z. Albert, and L. Vaidman, How the result of a measurement of a component of the spin of a spin-1/2 particle can turn out to be 100. Phys. Rev. Lett. {\bf 60}, 1351 (1988).

\bibitem{weak2} Y. Aharonov and L. Vaidman, Properties of a quantum system during the time interval between two measurements. Phys. Rev. A {\bf 41}, 11 (1990).

\bibitem{weak3} J. Dressel, M. Malik, F. M. Miatto, A. N. Jordan, and R. W. Boyd, {\it Colloquium:} Understanding quantum weak values: Basics and applications. Rev. Mod. Phys. {\bf 86}, 307 (2014).

\bibitem{S} G. Vallone and D. Dequal, Strong Measurements Give a Better Direct Measurement of the Quantum Wave Function. Phys. Rev. Lett. {\bf 116}, 040502 (2016).

\bibitem{Josza} R. Jozsa, Complex weak values in quantum measurement. Phys. Rev. A {\bf 76}, 044103 (2007).

\bibitem{app1} J. Z. Salvail, M. Agnew, A. S. Johnson, E. Bolduc, J. Leach, and R. W. Boyd, Full characterization of polarization states of light via direct measurement. Nat. Photonics {\bf 7}, 316 (2013).

\bibitem{app2} M. Malik, M. Mirhosseini, M. P. J. Lavery, J. Leach, M. J. Padgett, and R. W. Boyd, Direct measurement of a 27-dimensional orbital-angular-momentum state vector. Nat. Commun. {\bf 5}, 3115 (2014).

\bibitem{app3} S. Kocsis, B. Braverman, S. Ravets, M. J. Stevens, R. P. Mirin, L. K. Shalm, and A. M. Steinberg, Observing the Average Trajectories of Single Photons in a Two-Slit Interferometer. Science {\bf 332}, 1170 (2011).

\bibitem{app4} Z. Shi, M. Mirhosseini, J. Margiewicz, M. Malik, F. Rivera, and R. W. Boyd, Scan-free direct measurement of an extremely high-dimensional photonic state. Optica {\bf 2}, 388 (2015).

\bibitem{matter} T. Denkmayr, H. Geppert, H. Lemmel, M. Waegell, J. Dressel, Y. Hasegawa, and S. Sponar, Experimental Demonstration of Direct Path State Characterization by Strongly Measuring Weak Values in a Matter-Wave Interferometer. Phys. Rev. Lett. {\bf 118}, 010402 (2017).

\bibitem{holo1} J. W. Goodman and R. W. Lawrence, Digital Image Formation From Electronically Detected Holograms. Appl. Phys. Lett. {\bf 11}, 77 (1967).

\bibitem{holo2} I. Yamaguchi and T. Zhang, Phase-shifting digital holography. Opt. Lett. {\bf 22}, 1268 (1997).


\bibitem{compressive} M. Mirhosseini, O. S. Maga\~{n}a-Loaiza, S. M. H. Rafsanjani, and R. W. Boyd, Compressive Direct Measurement of the Quantum Wave Function. Phys. Rev. Lett. {\bf 113}, 090402 (2014).


\bibitem{Density matrix}  Luca Calderaro, Giulio Foletto, Daniele Dequal, Paolo Villoresi, Giuseppe Vallone, arXiv:1803.10703.

































\end{thebibliography}
\end{document}